\documentclass[12pt,paper,notoc]{JHEP3}

\def \beq {\begin{equation}}
\def \eeq {\end{equation}}

\usepackage{epsfig}

\title{Quasinormal mode
characterization of evaporating mini black holes}

\author{Elcio Abdalla, Cecilia B. M. H. Chirenti   \\
  Instituto de F\'isica,
Universidade de S\~ao Paulo,\\ C.P. 66318,   05315-970 - S\~ao Paulo, SP,
Brazil \\
E-mail: \email{eabdalla@fma.if.usp.br} \\
E-mail: \email{cecila@fma.if.usp.br} }
\author{Alberto Saa \\
  Departamento de Matem\'atica Aplicada, \\
IMECC -- UNICAMP, C.P. 6065, \\
13083-859 -  Campinas, SP, Brazil \\
E-mail: \email{asaa@ime.unicamp.br}}


\abstract{
According to recent theoretical developments, it might be possible to produce
mini black holes in the high energy experiments in the LHC at CERN.
We propose here a
model based on the $n$-dimensional Vaidya metric in double null coordinates for these decaying black
holes. The
associated    quasinormal modes are considered. It is shown that only in the very last instants of the
evaporation process
the stationary regime for the quasinormal modes is broken, implying   specific
power spectra for the perturbations around these mini black-holes.
From scattered   fields   one could recover, in principle, the black hole
parameters as well as the number of extra dimensions.
The still
mysterious final fate of such objects should not alter significantly our main conclusions.
}

\preprint{gr-qc/0703071}

\begin{document}

\section{Introduction}

New models with extra dimensions\cite{Arkani} predict the production of mini
black holes in particle accelerators with sufficient large energies.
Such events are
expected to be obtained in the LHC at CERN\cite{Dimopoulos}.
These $n$-dimensional mini black holes    are expected to be highly interacting, and,
once  formed, Hawking radiation\cite{Hawking1} is expected to settle   after possible transient stages.
Their phenomenological and observational
consequences   have been intensively discussed. (See \cite{conseq} for recent
reviews). In particular,
it is expected that their net radiated power, and hence, their mass decreasing
rate, be driven by the $n$-dimensional Stefan-Boltzmann law\cite{CDM}, leading to
\begin{equation}
\frac{d}{dt}\left(\frac{m}{M_{\rm P}} \right) = - \frac{a_n}{t_{\rm P}} \left(\frac{m}{M_{\rm P}}\right)^{-\frac{2}{n-3}} ,
  \label{dMdt}
\end{equation}
where $t_{\rm P}$ and $M_{\rm P}$ stand for  the Planck time and mass, respectively, while
 $a_n$   is the effective $n$-dimensional Stefan-Boltzmann constant\cite{CDM},
 which depends on the available emission channels for the   Hawking radiation\cite{Page}.
Typically, however, one should expect $a_n\approx 10^{-3}$.
Eq. (\ref{dMdt}) can be immediately integrated,
\begin{equation}
  m(t) = m_0\left(1 -   \frac{t}{t_{0}}\right)^{\frac{n-3}{n-1}},
  \label{real_mass}
\end{equation}
$0\le t\le t_0$,
where the lifetime $t_0$ of a black hole with initial mass $m_0$ is given by
\beq
t_0 = \frac{n-3}{n-1}\left(\frac{m_0}{M_{\rm P}}\right)^{\frac{n-1}{n-3}}\frac{t_{\rm P}}{a_n}.
\eeq
Following Arkani-Hamed {\em et al}\cite{Arkani}, the phenomenology of such
mini LHC black holes can be studied by setting  the Planck scale in order to
have $M_{\rm P} \approx 1$TeV.

We recall that (\ref{dMdt}) is not expected to be valid in the very
final stages of the black hole evaporation, where
the appearance of new emission channels for Hawking radiation can induce
changes\cite{Page} in the value of the constant $a_n$. Perhaps even
the usual adiabatic
derivation of Hawking radiations is not valid any longer\cite{adiabatic}. We do
not address these points here. We assume that the black hole evaporates
  obeying (\ref{real_mass})  for   $0\le t \le t_0$.
The numerical analysis, however, requires the introduction of a regularization for
the final instants of the evaporation process.
Nevertheless, as we will see,
our main results do not depend on such final details.

Here, we consider the quasinormal modes (QNM) associated to a radiating
$n$-dimensional black hole
with the decaying mass (\ref{real_mass}).
Since the preferred emission channels for Hawking radiation correspond to massless
fields, we model
these  evaporating mini black holes by means of an
$n$-dimensional
Vaidya metric  \cite{Vaidya,IV} in double-null coordinates\cite{WL,GS,Saa}. The Vaidya metric
corresponds to the  solution of
Einstein's equations with spherical
symmetry in the presence of a radial flow of unpolarized
radiation.
Such evaporating mini black holes, however, are not
expected to emit isotropically\cite{bulk,bulk1} and, hence,
any realistic
model should not be spherically symmetric.
Furthermore, a typical mini black hole created in the LHC environment should not have
zero angular momentum.
Our simple model, nevertheless, is a step toward the
construction of more realistic ones.  The identification of stationary regimes\cite{ACS} for the QNM
in the non-spherically symmetrical case,
for instance, could simplify the analysis of more realistic configurations.
We will return to these issues in the last section.

In $n$-dimensional double-null coordinates $(u,v,\theta_1,\dots,\theta_{n-2})$,
the Vaidya metric has the form\cite{IV}
\begin{equation}
  ds^{2} = -2g(u,v)du dv + r^{2}(u,v)d\Omega^{2}_{n-2},
  \label{metric}
\end{equation}
where $d\Omega^{2}_{n-2}$ stands for the unity $(n-2)$ dimensional sphere, and
$g(u,v)$ and $r(u,v)$ are smooth non vanishing functions
obeying\cite{Saa}
\begin{eqnarray}
  g  &=&  -\partial_{u}r ,
  \label{f(u,v)}\\
  \partial_{v}r  &=&  \frac{1}{2}-\frac{m(v)}{(n-3)r^{n-3}},
  \label{r(u,v)}
\end{eqnarray}
where,
for the present case of an outgoing radiation field ($m'<0$), $v$  corresponds
to the retarded time
coordinate. We adopt hereafter natural unities ($t_{\rm P} = M_{\rm P} = \ell_{\rm P} = 1$).

Our choice for the mass function $m(v)$ is guided
by the solution (\ref{real_mass}). Nevertheless,
the final stage of a black hole evaporation is a rather subtle issue.
A black hole could evaporate up to zero mass as described by (\ref{real_mass}),
leaving behind an empty
Minkowski-like spacetime\cite{mink} (or even a naked singularity\cite{naked}), or it could evaporate
until it reaches a minimum mass, that is, turning into a massive remnant\cite{remnant}.
In order to circumvent these problems in the numerical analysis, we   introduce a regularization
for the final stage of the evaporation process. We consider the   mass function
  \begin{eqnarray}
  m(v) &=& \left\{
  \begin{array}{l}
  m_0\left(1 -   \frac{v}{v_{0}}\right)^{\frac{n-3}{n-1}}, \quad 0 \le v < v_1 <   v_0, \\
    A - B\tanh \rho (v-v_1), \quad v > v_{1},
  \end{array}
  \right.
  \label{hyperbolic}
\end{eqnarray}
with $\rho>0$. The constants
 $A$, $B$ and $\rho$ are determined by
imposing conditions for the continuity of $m(v)$ and its first derivative
in $v = v_{1}$. Clearly,
 $A - B = m_{\rm F}  $  is the mass of the final remnant. The regularization
 is effective only at the final instants of the evaporation process, $(v_0-v_1)/v_0 \ll 1$,
 and $m_{\rm F} \ll 1$TeV. We will show
  that, during the major part of the evaporation process,
  the stationary regime for the quasinormal modes described in \cite{ACS} holds, implying a specific,
 and perhaps observable, power spectrum for the perturbations around these mini black-holes.
 Besides, the  perturbation  power spectra
 should not depend  significantly on the final fate of the black-hole.

\section{The Quasinormal modes}

We decompose a generic perturbation field $\phi$ as
\beq
\phi = \sum_{\ell m} r^{-\frac{n-2}{2}} \psi_\ell(u,v) Y_{\ell m}(\theta_1,\dots,\theta_{n-2}),
\eeq
where $Y_{\ell m}$ stands for the scalar spherical harmonics on the $(n-2)$ unity sphere, for
which $\partial_{\Omega_{n-2}}^2 Y_{\ell m} = -\ell(\ell + n-3) Y_{\ell m}$,
where $\ell = 0,1,2,\dots$,
and $m$ denotes a set of $(n-3)$ integers $(m_1,m_2,\dots,m_{n-3})$ satisfying
$\ell \ge m_{n-3} \ge m_2 \ge |m_1|$ (See \cite{CM} for a concise description of higher dimensional
spherical harmonics). By using   (\ref{f(u,v)}) and (\ref{r(u,v)}),
the equation for $\psi_\ell$ reads
\begin{equation}
\frac{\partial^{2}\psi_\ell}{\partial u\partial v}  +
    g(u,v) V( u,v )   \psi_\ell = 0,
\label{scalar}
\end{equation}
where
\begin{equation}
V(u,v) = \frac{1}{2} \left( \frac{\ell(\ell +n-3)}{r^2} + \frac{(n-2)(n-4)}{4r^2} +
\frac{(1-\sigma^2)(n-2)^2}{4r^{n-1}}m(v)\right).
\end{equation}
The constant $\sigma$ determines the type of the perturbation considered: $\sigma =0$ corresponds to
scalar and gravitational tensor perturbations, $\sigma =2$ to gravitational vector perturbations,
$\sigma =2/(n-2)$ to electromagnetic vector perturbations, and $\sigma =2-2/(n-2)$, finally, to
electromagnetic scalar perturbations\cite{CHM}.

We perform an exhaustive numerical analysis of the equations (\ref{f(u,v)}), (\ref{r(u,v)}), and
(\ref{scalar}) along the same lines of the method proposed in \cite{ACS}. In particular, we could
verify that the QNM stationary behavior for slowly varying masses reported in \cite{ACS}
is not altered in higher dimensional spacetimes, see Fig. \ref{fig1}.
Hence,
\FIGURE[ht]{\epsfig{file=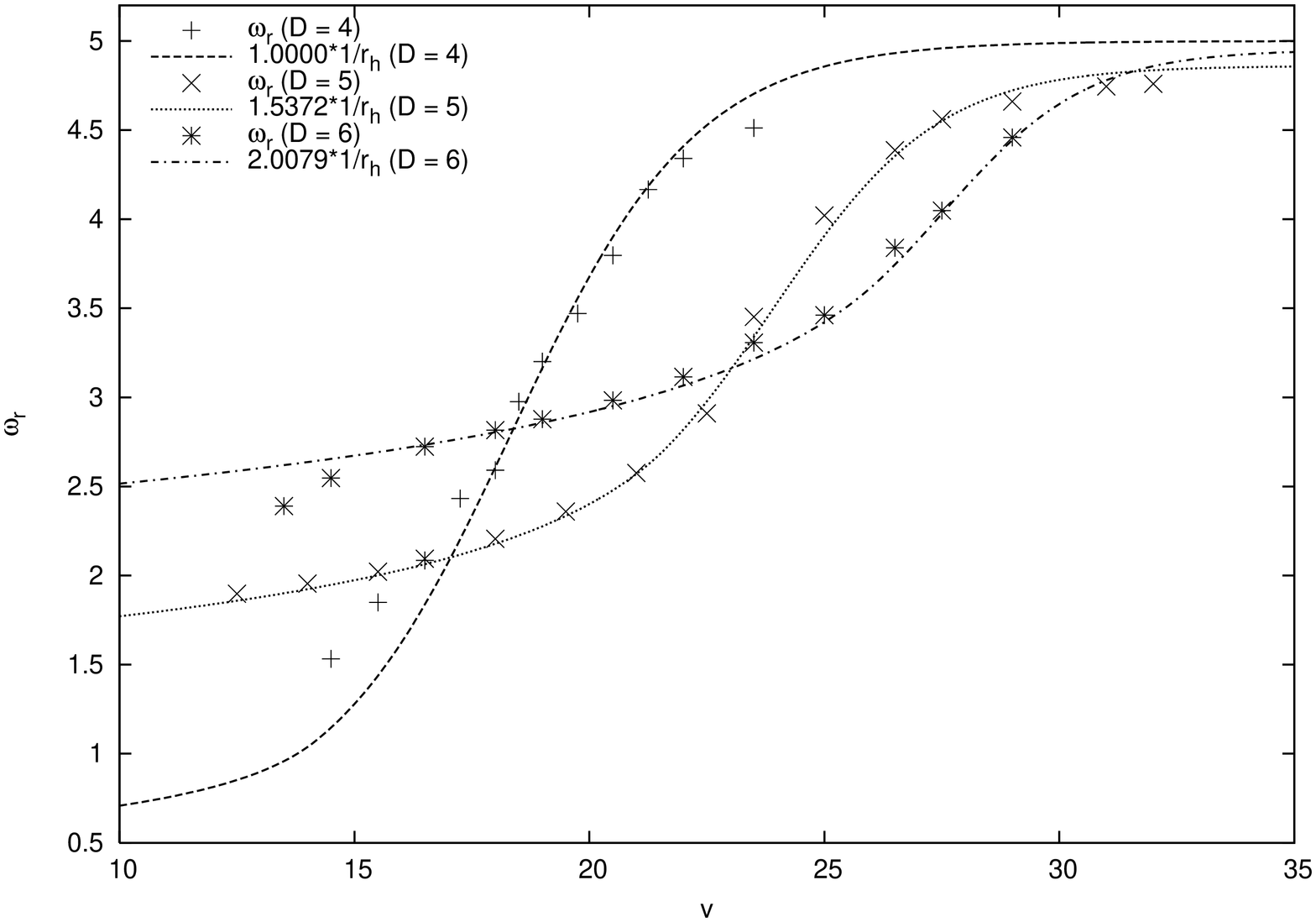,angle=-90,width=10cm}\caption{QNM ``instantaneous" frequencies (see \cite{ACS}) for the equation (\ref{scalar}).
In the stationary regime, the frequencies
follow the track corresponding to $1/r_h(v)$. The cases depicted here corresponds to
$\sigma=0$,
$\ell =2$,  and $a=0.02$. Note that, the smaller the value of $a_n$, the better   the fit. \label{fig1}}}
 provided the mass function $m(v)$ varies slowly, the QNM
of (\ref{scalar}) set down in a stationary regime, and the associated frequencies
($\tilde{\omega}_{\rm R}$)
and damping terms ($\tilde{\omega}_{\rm I}$)  follow the track corresponding to $1/r_h(v)$, where
$r_h$ is the (aparent) horizon\cite{GS} radius of an $n$-dimensional black-hole of mass $m(v)$. In a more quantitative
way, one has here
\begin{equation}
\label{freq}
\frac{\tilde{\omega}_{\rm R,I}(v)}{\omega_{\rm R,I}}  =   \frac{r_h(0)}{r_h(v)} =
\left( 1 -   \frac{v}{v_{0}}\right)^{-\frac{1}{n-1}},
\end{equation}
where ${\omega}_{\rm R,I}$ stand for the oscillation frequency (${}_{\rm R}$) and damping term
(${}_{\rm I}$)
 of the QNM corresponding to an $n$-dimensional Schwarzschild black hole with mass $m_0=m(0)$.
We notice that
 the  relation $\omega_{\rm R} \propto 1/r_h$
  for $n$-dimensional Schwarzschild black holes has been previously obtained
  by Konoplya in \cite{Konoplya}.
As in \cite{ACS},
we have used Gaussian initial conditions for all the analysis, although equivalent results
can be obtained for any localized initial condition.

Our simulations strongly indicate that
the condition for the QNM stationary regime\cite{ACS} must be generalized for
the case of $n$-dimensional black holes
as $|r_h''(v)| < |\tilde\omega_{\rm I}(v)|$,
where $\tilde\omega_{\rm I}(v)$ is the smallest damping term of the system.
In the present case, it reads
\beq
\left(1-\frac{v}{v_0} \right)^{2\frac{n-2}{n-1}} > a_n^2\left[\frac{n-2}{(n-3)^2}\left(\frac{2}{n-3} \right)^{\frac{1}{n-3}}\frac{m_0^{-\frac{2n-3}{n-3}}}{\omega_{\rm I}}\right]
\eeq
where $\omega_{\rm I}$ is the smallest   damping term
of an $n$-dimensional Schwarzschild black hole with mass $m_0=m(0)$, typically corresponding to
scalar
perturbations.
For the LHC mini black holes,
the term between square brackets should be of order of unity,
  irrespective of $n$. Thus,
only in the very late times of the evaporation process (for typical small values of $a_n$,
for less than the last $a_n^{(n-1)/(n-2)}$ fraction of the lifetime period) the stationary regime is broken.
Hence,
the late time exponentially suppressed perturbation
of (\ref{scalar}) is well approximated by
\beq
\label{pert}
\tilde\psi(v) = e^{-{\tilde\omega_{\rm I}}v}\sin\left(\tilde\omega_{\rm R}v +\delta \right)
\eeq
for $0\le v < v_0$, and $\tilde\psi(v) = 0$ for $v \ge v_0$,
where $\tilde\omega_{\rm R,I}$   are themselves functions of $v$ given by
(\ref{freq}), and $\delta$ is an arbitrary phase.

Our main observation is that, for the typical values of the parameter $a_n$ and
$m_0\approx$ 1TeV,
the Fourier spectrum $\tilde\Psi(f)$ of the stationary
perturbations (\ref{pert}) is  very
close to the the Fourier spectrum $\Psi(f)$ of the
perturbations corresponding to a
$n$-dimensional Schwarzchild case ($m(v)=m_0$),
\beq
\label{pert1}
\psi(v) = e^{-{\omega_{\rm I}}v}\sin\left(\omega_{\rm R}v +\delta \right)
\eeq
for $v \ge 0$.
This fact, clearly illustrated in Fig. \ref{fig2}, certainly would
\FIGURE[ht]{\epsfig{file=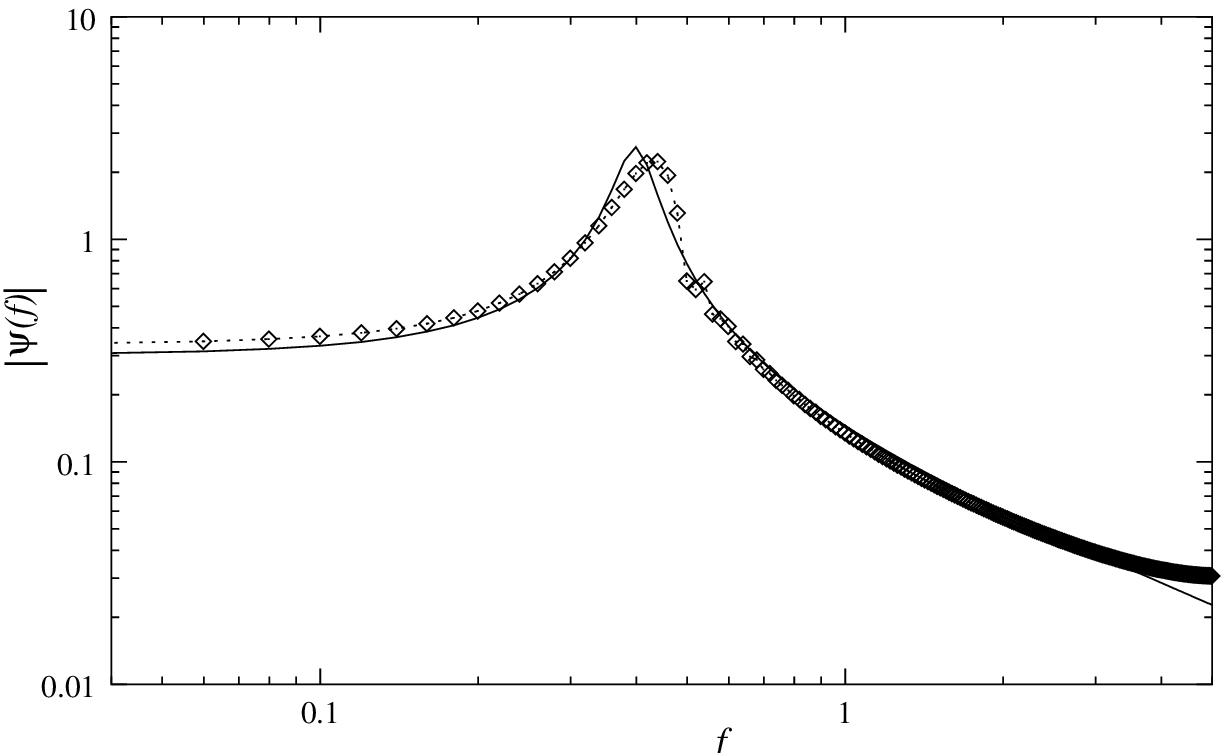,width=10cm}\caption{
 The  power spectra: $|\Psi(f)|$, the solid line, given by Eq. (\ref{power}); and
$|\tilde\Psi(f)|$, the line with points, calculated numerically from (\ref{pert}) and
(\ref{freq}).
They are indeed very close for the typical values of $a$ and
$m_0\approx$ 1TeV.
In particular, both spectra exhibit similar  pronounced peaks.
Note that the discrepancies for large values of
$\omega$ are due to the FFT aliasing effect for frequencies larger than the Nyquist critical
frequency\cite{NR}, and not to real differences between $|\Psi(f)|$ and
$|\tilde\Psi(f)|$. The case depicted here corresponds to   $n=4$, $a_4=0.002$,
$\omega_{\rm R}=0.25$, $\omega_{\rm I}=0.01$,  and $\delta = 0$.
\label{fig2}}}
 deserve  a more rigorous analysis. Some simple estimations, however, do  endorse the
 observation. From the linearity of the Fourier transform and Parseval's theorem, we have
 that
 \beq
 \label{parseval}
 \int_0^\infty\left( \psi(v) - \tilde\psi(v) \right)^2 dv =
 \int^\infty_{-\infty}\left| \Psi(f) - \tilde\Psi(f)\right|^2 df.
 \eeq
Hence, provided the left handed side of (\ref{parseval}) be small,
$\Psi(f)$ will be close (in the $L^2$ norm) to $\tilde\Psi(f)$.
The integral in the left handed side of (\ref{parseval}) can be split as
\beq
 I_1  + I_2 =
 \int_0^{v_2}\left( \psi - \tilde\psi \right)^2 dv
 + \int_{v_2}^{\infty}\left( \psi - \tilde\psi \right)^2 dv.
\eeq
The second integral can be estimated as
\begin{equation}
I_2  \le
2\left(\int_{v_2}^{\infty}  \psi^2 dv  +
\int_{v_2}^{\infty}  \tilde\psi^2 dv \right) \le 4  \int_{v_2}^{\infty} e^{-2{\omega_{\rm I}}v} dv
= 2\frac{e^{- 2{\omega_{\rm I}}v_2}}{{{\omega_{\rm I}}}}.
\end{equation}
Typically, $\omega_{\rm R}$ and $\omega_{\rm I}$  are of the same order (unity), while
$a_n$ is much smaller ($10^{-3}$).
If one chooses $v_2$ corresponding, for instance, to  10 oscillation
cycles of $\psi(v)$, the value of $I_2$ will be less than $e^{-20}$. This is the
error involved in approximating the left handed side of (\ref{parseval}) by $I_1$.
On the other hand, during the 10 first oscillation
cycles of $\psi(v)$, the variations of $\tilde\omega_{\rm R,I}(v)$ are of the
order   $1-(100/99)^{1/(n-1)}$ for black holes with initial mass $m_0$ = 1 TeV.
Hence, in the interval $[0,v_2]$, $\psi(v)$
is indeed very close to $\tilde\psi(v)$, implying that the left handed side of (\ref{parseval})
is small and, finally, that  $\tilde\Psi(f)$ is close to $\Psi(f)$.

The Fourier spectrum $\Psi(f)$ of the
 perturbation   (\ref{pert1}) can be easily calculated
\begin{equation}
\Psi(f) = \frac{1}{\sqrt{2\pi}} \int_0^\infty \psi(t) e^{-if t}\, dt
= \frac{1}{\sqrt{2\pi}}
\frac{\omega_{\rm R}\cos\delta +\left(\omega_{\rm I} +if\right)\sin\delta}{\left(\omega_{\rm R}\right)^2 +
\left(if +\omega_{\rm I} \right)^2}.
\end{equation}
The associated power spectrum
\beq
\label{power}
\left|\Psi(f)  \right| = \frac{1}{\sqrt{2\pi}} \sqrt{\frac{(\omega_{\rm R}\cos\delta + \omega_{\rm I}\sin\delta)^2+f^2\sin^2\delta}
{(\omega_{\rm R}^2 + \omega_{\rm I}^2 - f^2)^2 + 4f^2\omega_{\rm I}^2}}
\eeq
has a pronounced peak (see Fig. \ref{fig2}) at $f_{max}$ given by
\begin{equation}
\label{g}
\frac{f^2_{max} -\left(\omega_{\rm R}^2 - \omega_{\rm I}^2 \right)}
{\left(\omega_{\rm R}^2+ \omega_{\rm I}^2 \right)^2 - f^4_{max} } = g(f^2_{max})
=\frac{1}{2} \left(
\frac{\sin\delta}{\omega_{\rm R}\cos\delta + \omega_{\rm I}\sin\delta}
\right)^2,
\end{equation}
from where we can conclude that
 \beq
 \label{range}
 \sqrt{\omega_{\rm R}^2- \omega_{\rm I}^2} \le f_{\rm max} \le \sqrt{\omega_{\rm R}^2+ \omega_{\rm I}^2},
 \eeq
 provided $|\omega_{\rm R}| > |\omega_{\rm I}|$, see
Fig. \ref{fig3}.
\FIGURE[ht]{\epsfig{file=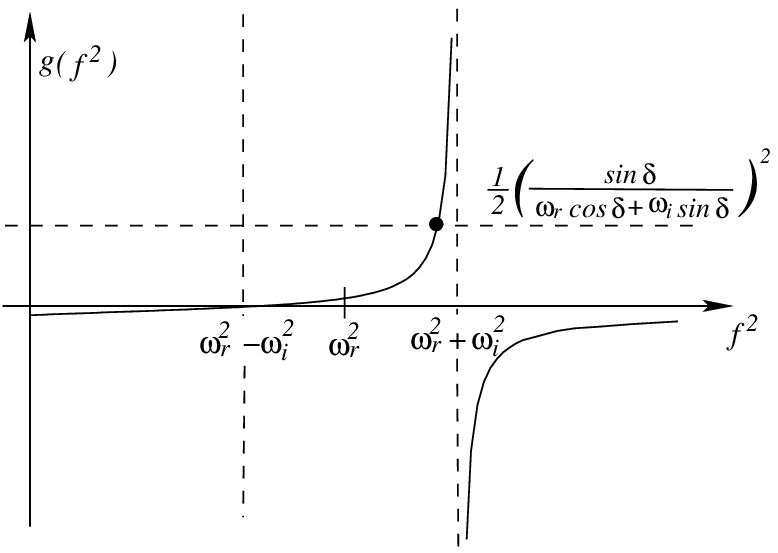,width=10cm}\caption{
Graphical solution of (\ref{g}). It is clear that, for arbitrary phases $\delta$,
the peak of the power spectrum (\ref{power}) is located in the range given by (\ref{range}).
\label{fig3}
}}

\section{Final Remarks}

The most interesting part of our results concerns the characterization of
the signals that might come out of the black hole probe. From the peaks in the power
spectra of the perturbations around the evaporating mini black holes, it is possible to
determine $\omega_{\rm R}$ and $\omega_{\rm I}$ and, consequently, infer
some of the black hole parameters as its initial mass $m_0$ and even the dimension $n$ of the
spacetime where it effectively lives. We do not expect, of course,  the
gravitational perturbations associated to these mini black hole to be measurable.
However, we remind that the QNM analysis can be applied for any   test
field propagating around the black hole.
In particular, it also applies for real (confined to the brane) electromagnetic
perturbations. Moreover, since
 these higher-dimensional mini-black holes are expected to emit large fractions of their
masses in the bulk\cite{bulk}, the assumption of a spherically symmetrical $n$-dimensional Vaidya metric
can be justified as a first approximation.

According to the discussion of last section,
the late time behavior of  electromagnetic waves scattered
by these evaporating mini  black holes should exhibit a power spectrum as that
one depicted in Fig. \ref{fig2}, since the electromagnetic perturbations will be
also of the form (\ref{pert}) for large times. Furthermore,   the energy carried
by these perturbations is negligible and, thus, they are expected to
 be fully sensitive to the higher dimensional
dynamics of the black hole. Hence,
from a precise determination
of the peak location for real electromagnetic perturbations,
we can get the  relevant parameters of the mini black hole,
including the number of extra dimensions.
For instance, for   4-dimensional
black holes, the QNM frequencies and damping terms for the first
electromagnetic perturbations\cite{Iyer} are
$\omega_{\rm R}=0.2483$ and
$\omega_{\rm I}=0.0925$, implying that
  the frequency peak of the electromagnetic power spectrum be in the range
\beq
 \left( \frac{m_0}{1\rm TeV} \right) {\hbar f} =  230 \ {\rm to}\  265 \ \rm GeV.
\eeq
Typically, the larger is the number of extra dimensions, the larger will be the
peak frequency, even surpassing the limit of 1 TeV. However,
one can, in principle, determine the relevant parameters of the mini black hole
by comparing the peaks of power spectra of   real (confined to the brane) perturbation fields
with the calculated higher-dimensional QNM frequencies and damping terms for the
 $n$-dimensional
Schwarzschild black-hole\cite{KonoZhi}.

\acknowledgments

This work has been supported by Funda\c c\~ao
de Amparo \`a Pesquisa do Estado de S\~ao Paulo {  (FAPESP)} and Conselho
Nacional de Desenvolvimento Cient\'\i fico e Tecnol\'ogico {  (CNPq)},
Brazil.

\end{document}